\documentclass[aps,pra,twocolumn,superscriptaddress]{revtex4-1}
\usepackage{bm}
\usepackage{color} 
\usepackage{graphicx}
\usepackage{epsfig}
\usepackage{hyperref}
\usepackage{natbib}
\usepackage{amsmath}
\usepackage{amssymb}
\usepackage{bbm}
 
\newcommand{\rmi}{{\rm i}}
\newcommand{\rmd}{{\rm d}}
\newcommand {\e}{{\rm e}}

\graphicspath{{./},{./figs/}}

\begin{document}

\title{Doppler--Raman crossover in resonant scattering}

\author{\firstname{A.~V.} \surname{Poshakinskiy}}
\email{poshakinskiy@mail.ioffe.ru}
\affiliation{Ioffe  Institute, St.~Petersburg 194021, Russia}
\author{\firstname{A.~N.} \surname{Poddubny}}
\affiliation{Ioffe  Institute, St.~Petersburg 194021, Russia}
\author{\firstname{N.~A.} \surname{Gippius}}
\affiliation{Skolkovo Institute of Science and Technology, Skolkovo 143025, Russia}


\begin{abstract}
We consider theoretically light scattering by a resonant layer that periodically moves in real space. At small frequencies of motion the scattered light spectrum reveals the frequency shift that is governed by the Doppler effect. At higher motion frequencies,  the scattered light spectra acquire sidebands stemming from the Raman effect. We investigate the crossover between these two regimes and propose a realistic quantum well structure for its observation.
\end{abstract}

\maketitle


\section{Introduction}

Frequency conversion is a basic nonlinear optical process, widely used for signal multiplexion and frequency comb generation~\cite{ boyd2003nonlinear}. An alternative approach is offered by optomechanical and optoacoustic nonlinearities  \cite{Fan2016,Preble2007,scruby_laser_1990}. The advantage of optomechanical systems is their relative compactness,  enabling manipulation of spectrum of classical and quantum light on the nanoscale~\cite{Kuramochi2016,Demenev2019}. Specifically, mechanical motion or deformation leads to a dynamical modification of optical properties such as refractive index~\cite{Farmer2014}, resonance frequency \cite{Jusserand2015} and optical gain~\cite{Brggemann2011}, allowing one to control light intensity and frequency~\cite{Berstermann2009,Berstermann2012},  as well as propagation direction~\cite{Poshakinskiy2019}. 
 The simplest example of motion-induced frequency conversion is given by the Doppler effect. The motion of the light source with respect to the observer with the constant velocity $v$ leads to the shift of the observed light frequency by $(v/c)\omega_0$, where $\omega_0$ is the emitter frequency and $c$ is the speed of light. Another opportunity for frequency conversion is given by the Raman effect. When the optical properties of the medium oscillate at frequency $\Omega$, the scattered light spectrum acquires sidebands shifted by $\Omega$ from the initial frequency. Both Doppler and Raman effects can be realized if one considers light scattering on an object periodically moving in space. 
While the two effects share the same origin, the corresponding frequency shifts are distinct.  The Raman shift of the scattered light frequency is always a multiple of mechanical motion frequency $\Omega$, while the Doppler frequency shift $(\Omega u_0/c)\omega_0 $, being proportional to the motion velocity, scales linearly with the mechanical motion amplitude $u_0$.  In this paper, we consider light scattered by a resonant layer oscillating in space. We calculate the scattered light spectra, identify the regimes where Doppler and Raman shifts can be observed, and investigate the crossover between them.

\section{Model}

\begin{figure}[tb]
  \includegraphics[width=.6\columnwidth]{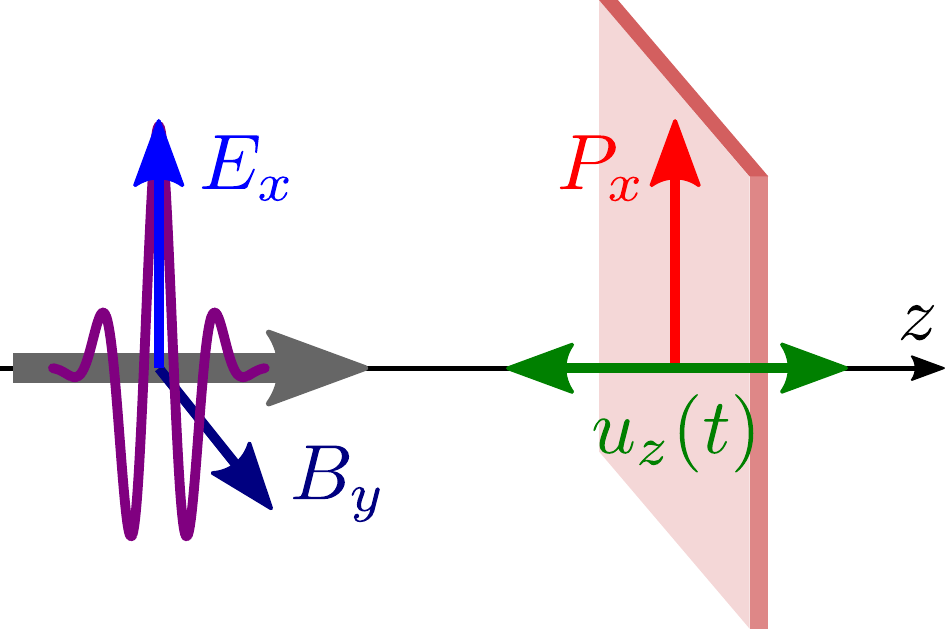}
\caption{A sketch of a short light pulse incident upon a layer periodically moving in space.}\label{fig:0}
\end{figure}

We study light reflection and transmission through a moving layer with the resonant polarizability. We suppose that light is incident along the layer normal $z$ and linearly polarized along $x$, see Fig.~\ref{fig:0}. The layer motion  along $z$ axis is characterized by the displacement $u_z(t)$. Such system can be described by the action
\begin{align}\label{eq:S}
S &=  \frac{1}{8\pi} \int \left[E_x^2 (z,t)- B_y^2(z,t)\right] dz\, dt \\\nonumber
 &+ \int P_x(t) \left\{ E_x[u_z(t),t] + \frac{\dot{u}_z(t)}{c} B_y[u_z(t),t] \right\} dt + S_{d} \,,
\end{align}
where dot denotes the time derivative. 
The first term in Eq.~\eqref{eq:S} is the action of free electromagnetic field where the electric and magnetic fields can be be expressed via vector potential $A_x(z,t)$ as $E_x = -(1/c)(\partial A_x /\partial t)$ and $B_y = \partial A_x/\partial z$.
The second term  in Eq.~\eqref{eq:S} describes interaction of the layer polarization $P_x(t)$ with electromagnetic field, see Appendix~\ref{app:L} for the derivation. The last term $S_d$ is the action describing layer polarization. We are interested in the layer with resonant response, so we take the simplest model of a harmonic oscillator with the eigenfrequency $\omega_0$, described by the Lagrangian
\begin{align}
S_d = \frac{\pi}{2c\Gamma_0} \int \left\{ \left( \frac{d P_x}{dt_0}\right)^2 - \omega_0^2 P_x^2 \right\} dt_0 \,.
\end{align}
Here $dt_0 = dt\sqrt{1-\dot{u}_z^2(t)}$ is the time in the moving reference frame and the constant $\Gamma_0$  is the radiative decay rate of the oscillator, as justified below. 

The action Eq.~\eqref{eq:S} yields the Euler--Lagrange dynamic equations
\begin{align}
&\frac1{c^2}\frac{\partial^2 A_x}{\partial t^2} - \frac{\partial A_x}{\partial z^2 } = \frac{4\pi}c \frac{d P_x}{dt} \delta [z-u_z(t)] \label{eq:A}\,, \\
&\frac{d^2 P_x}{dt_0^2}+ 2\Gamma\frac{d P_x}{dt_0} + \omega_0^2 P_x = - \frac{\Gamma_0}{\pi} \, \frac{d}{dt_0}  A_x[u_z(t),t] \label{eq:P}\,,
\end{align}
where we introduced in Eq.~\eqref{eq:P} the oscillator damping with the rate $\Gamma$ accounting for non-radiative decay processes. 
Equation~\eqref{eq:A} can be solved using the Green function $G(z,t) = (c/2) \, \theta(ct - |z|)$ and yields
\begin{align}\label{eq:Ad}
&A_x(z,t) = A_x^{(0)}(z,t) + 2\pi P_x[t^*(z,t)],  
\end{align}
where $A_x^{(0)}$ is the vector potential of the incident wave and $t^*(z,t) = t-|z-u_z[t^*(z,t)]|/c$.  In particular, $t^*[u_z(t),t] = t$, so $A[u_z(t),t] = A_x^{(0)}[u_z(t),t] +2\pi P_x(t)$. Substituting this into Eq.~\eqref{eq:P} we obtain a closed-form equation for the polarization
\begin{align}\label{eq:evd}
\frac{d^2 P_x}{dt_0^2} + 2\Gamma'\frac{d P_x}{dt_0}+ \omega_0^2 P_x = - \frac{\Gamma_0}{\pi} \, \frac{d}{dt_0}  A_x^{(0)}[u_z(t),t] \,,
\end{align}
where $\Gamma' = \Gamma + \Gamma_0$. As follows from Eq.~\eqref{eq:evd}, interaction of the layer polarization with the electromagnetic field leads to the increase of the polarization decay rate by $\Gamma_0$. Therefore, the latter describes the radiative decay rate of the oscillator.  

\begin{figure*}[tb]
  \includegraphics[width=.99\textwidth]{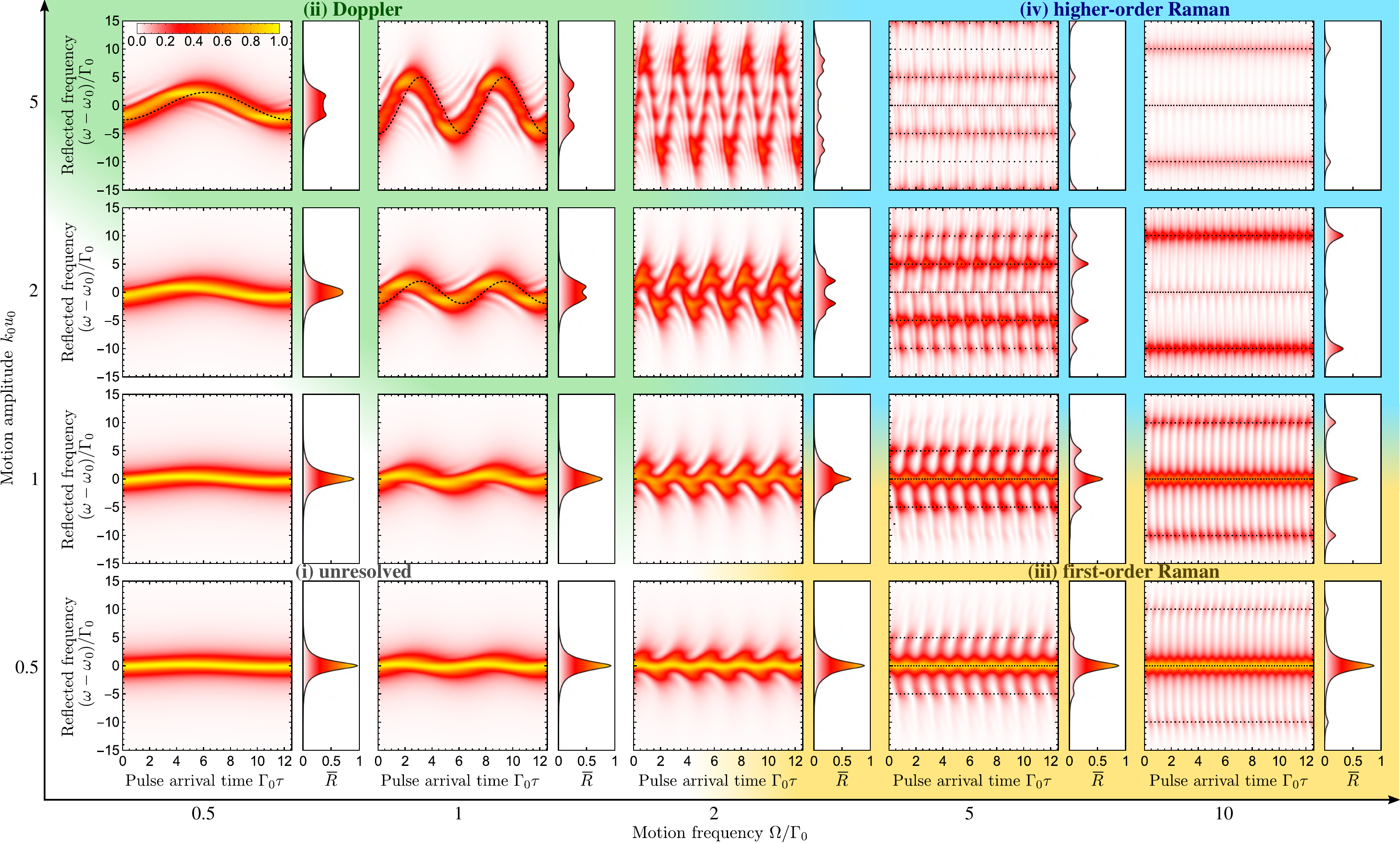}
\caption{The reflected light spectra after excitation of the resonant layer periodically moving in space with a short pulse. 
The plots are calculated for various motion frequencies and amplitudes, revealing the crossover between Doppler and Raman effects (see colored areas). Left panels show the color plots of the spectra as a function of pulse arrival time $\tau$. Right panels show the $\tau$-averaged spectra. The non-radiative resonance broadening is supposed to be absent, $\Gamma=0$. }\label{fig:cros}
\end{figure*}

\begin{figure*}[tb]
  \includegraphics[width=.99\textwidth]{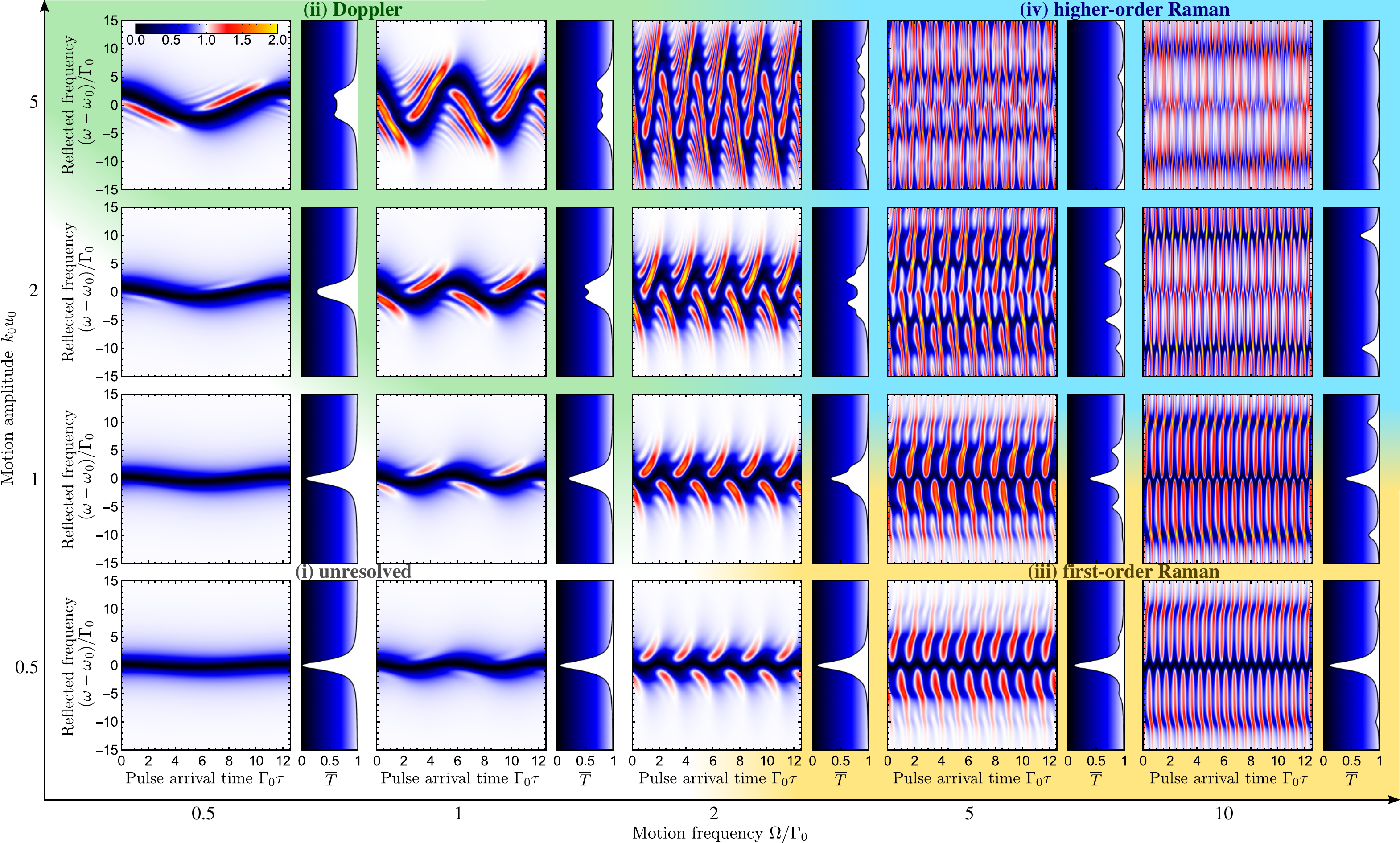}
\caption{The transmitted light spectra after excitation of the resonant layer periodically moving in space with a  short pulse. 
Left panels show the color plots of the spectra depending on the pulse arrival time $\tau$. Right panels show the $\tau$-averaged spectra. The non-radiative resonance broadening is supposed to be absent, $\Gamma=0$.
}\label{fig:crosT}
\end{figure*}

\section{Time-resolved spectroscopy}

In order to reveal the frequency conversion for light reflected from the oscillating layer and analyze the  crossover between Doppler and Raman effects, we propose to use optical spectroscopy with temporal resolution. In such approach, which is alternative to traditional frequency-resolved spectroscopy, one studies the response of the system to short optical pulses. When the probe pulse duration is smaller than the inverse width of the resonance, the reflected pulse spectrum matches the frequency dependence of the reflection coefficient~\cite{Andreani1998,Ammerlahn2000,Poshakinskiy2012}.
Here, since the system under study changes with time, the reflected pulse spectrum depends on the time of incident pulse arrival and characterizes the system reflectivity at that moment.  

We consider the illumination with a $\delta$-pulse, described by the electric field
\begin{align}
E^{(0)}(z,t) = \delta(t-\tau-z/c) \,
\end{align}
and  vector potential $A^{(0)}(z,t) = -c\theta(t-\tau -z/c)$, where $\tau$ is the time of pulse arrival to $z=0$.  In what follows we suppose that  the layer oscillation frequency $\Omega$ and the Doppler shift $ \omega_0\Omega u/c $ are much smaller than  the resonance  frequency $\omega_0$. Then, Eq.~\eqref{eq:evd} can be solved as 
\begin{align}
P_x(t) =- \frac{\Gamma_0}{\pi}\, \int   \frac{\e^{-\rmi \omega [t-\tau - u_z(\tau)/c]}}{\omega^2-\omega_0^2 + 2\rmi\omega\Gamma'}\, \frac{d\omega}{2\pi} \,,
\end{align}
Substituting this result into Eq.~\eqref{eq:Ad} we obtain the vector potential of the reflected wave, 
\begin{align}
&A_x^{(r)}(z,t) =\frac{2\Gamma_0}{\omega_0}\theta(\tau')\,  \sin \omega_0\tau' \,\e^{-\Gamma' \tau'},
\end{align}
where $\tau' = t+z/c-\tau-[u_z(\tau)+u_z( t+z/c)]/c $.  
The reflected pulse spectrum, $E_x^{(r)}(\omega) = \int E_x^{(r)}(\tau+t) \e^{\rmi\omega t} \, dt $, assumes the form 
\begin{align}\label{eq:r}
E_x^{(r)}(\omega)= -\Gamma_0 \int\limits_0^{\infty} \e^{\rmi (\omega-\omega_0+\rmi\Gamma')t +\rmi k_0[u_z(\tau)+u(\tau+t)]}\, dt \,,
\end{align}
where $k_0 = \omega_0/c$ and we assumed additionally that $\Gamma \ll \omega_0$. 
Similar consideration for the transmitted wave gives the spectrum
\begin{align}\label{eq:t}
E_x^{(t)}(\omega) &= 1 -\Gamma_0 \int\limits_0^{\infty} \e^{\rmi (\omega-\omega_0+\rmi\Gamma')t +\rmi k_0[u_z(\tau)-u(\tau+t)]}\, dt ,
\end{align}
where the unity term stems from the spectrum of the incident light pulse. The reflected and transmitted power spectra read $R(\omega) = |E_x^{(r)}(\omega)|^2$ and $T(\omega) = |E_x^{(t)}(\omega)|^2$. 

It is instructive to consider also the reflection and transmission spectra averaged over the probe arrival time $\tau$. Performing the averaging in Eqs.~\eqref{eq:r} and~\eqref{eq:t} we get
\begin{align}
&\overline R (\omega) = \frac{\Gamma_0^2}{\Gamma'} C(\omega-\omega_0)\:, \label{eq:aR} \\ \label{eq:aT}
&\overline T (\omega) = 1 - \frac{\Gamma_0(\Gamma_0+2\Gamma)}{\Gamma'} C(\omega_0-\omega)\:,
\end{align}
where 
\begin{align}
& C(\omega) = \text{Re} \int_0^\infty C(t) \, \e^{\rmi \omega t - \Gamma' t} dt \,, \label{eq:Cw} \\ \label{eq:Ct}
& C(t) = \langle \e^{ \rmi k_0 [u_z(\tau +t) - u_z(\tau)] }  \rangle_\tau\,,
\end{align}
the angular brackets denote averaging over $\tau$,
and we recall that $\Gamma' = \Gamma + \Gamma_0$. In the absence of non-radiative decay, $\Gamma=0$, the conservation law
\begin{align}
\overline R(\omega) + \overline T(2\omega_0-\omega) =1
\end{align}
is fulfilled, i.e., the sum of the reflected spectral power  taken at the frequency shifted up by Doppler effect and transmitted spectral power  taken at the frequency  shifted down, or vice versa, is equal to the spectral power of the incident pulse. If the motion of the layer is invariant under time-reversal symmetry, $C(t) = C(-t)$, then $C(t)$ is real and $C(\omega) = C(-\omega)$. In such case, both the reflected and transmitted power spectra are symmetric with respect to the resonance frequency $\omega_0$, while the conservation law assumes a simpler form  $\overline R(\omega) + \overline T(\omega) =1$. 
The total reflected power $\int \overline{R} (\omega)\, \rmd\omega/(2\pi) =\Gamma_0^2/\Gamma'$ does not depend on the specifics of layer motion, that only redistributes the power over the spectrum. 
\subsection{Periodic layer motion}
First, we consider the case of the periodic layer motion described by $u_z(t) = u_0 \sin \Omega t$. 
Shown in the Fig.~\ref{fig:cros} are the color plots of the reflected light spectra $R(\omega) $ depending on the  pulse arrival time $\tau$ and calculated for various motion amplitudes and frequencies. The graphs on the right present the $\tau$-averaged spectra $\overline R(\omega)$. 
We distinguish several regimes indicated by the color fill. 

When $\Omega \ll \Gamma'$, the pulse is reflected at the time scale smaller than motion period. Therefore, its spectral conversion is governed by the Doppler effect in the corresponding time moment. If the Doppler shift is small, $k_0 u_0 \Omega  \ll \Gamma'$ [regime (i), white area], the effect is smeared by resonance broadening. If $k_0 u_0 \Omega \gg \Gamma'$ [regime (ii), green area], the maximum of the reflected spectrum shifts strongly with the pulse arrival time $\tau$ following  the layer velocity, $\omega_0[1- (\Omega u_0/c) \cos \Omega t]$, indicated by the dashed curve, with a retardation on the order of $1/\Gamma'$. 


As the motion frequency  is increased, $\Omega \sim \Gamma'$, the sidebands start to appear around the central peak that still follows the cosine dependence of  the Doppler shift~\cite{Berstermann2009}.  The  case $\Omega \gg \Gamma'$ corresponds to the well-resolved Raman sidebands while the cosine dependence of the Doppler shift is fragmentized. In such regime, the $\tau$-averaged spectra $\overline R(\omega)$ is conveniently presented as  sum of Lorentzians at the sideband frequencies,
\begin{align}\label{eq:Jn2}
\overline R(\omega) = \sum_{n=-\infty}^{\infty } J_n^2(k_0 u_0)\,  \frac{\Gamma_0^2}{(\omega-n\Omega)^2+\Gamma'^2}, 
\end{align}
where $J_n$ is the Bessel function of the first kind.
 When the Raman scattering probability is weak, $k_0 u_0 \ll 1$, [regime (iii), yellow area], only the first order scattering sideband is revealed, see dashed lines indicating the sidebands at $\omega_0 \pm \Omega$. If $k_0 u_0 \gtrsim 1$  [regime (iv), yellow area], the sidebands of higher order begin to appear, while the amplitude at the initial frequency $\omega_0$ gets suppressed, in accordance with Eq.~\eqref{eq:Jn2}.  

Figure~\ref{fig:crosT} shows the color plots for the transmitted light spectra. As follows from Eq.~\eqref{eq:t}, the transmitted spectrum is the result of interference between the initial pulse (first term) and the scattered light (second term). Depending on the phase of the scattered light, the interference can be both constructive and destructive. As a result, the spectral intensity of the transmitted pulse can be both greater or smaller than that of the incident pulse~\cite{Berstermann2009}, which is encoded by red and blue colors in  Fig.~\ref{fig:crosT}, respectively.

\subsection{Stochastic layer motion}

Now we turn to the case when the layer displacement $u_z(t)$ is a stochastic function of time. The averaged reflection and transmission spectra are still given by Eqs.~\eqref{eq:aR}-\eqref{eq:Ct}, where the averaging over $\tau$ should be replaced by the averaging over the realizations of the function $u_z(t)$.  

We assume that the   displacement $u(t)$ has a Gaussian probability distribution. 
To evaluate the spectra, we rewrite Eq.~\eqref{eq:Ct} as $C(t) = \langle\e^{ \rmi (\omega_0/c)\int_0^t \dot{u}_z(t')dt'}\rangle$. Next, we expand the exponent into the series, use the Wick's theorem and obtain
\begin{align}\label{eq:C1}
C(t) = \e^{ (k_0^2/2) \int_0^t  \int_0^t \ddot{K}(t'-t'') \,dt' dt'' }\,,
\end{align}
where $K(t) = \langle u_z(0) u_z(t)\rangle$ is the correlation function of the layer displacement. 
Performing integration in Eq.~\eqref{eq:C1} we finally get
\begin{align}
C(t) = \e^{ k_0^2 \left[ K(t) - K(0) \right] }.  
\end{align}

As an example, we consider the case when the layer motion corresponds to a harmonic oscillator driven by a stochastic force with a white noise spectrum. The Fourier component of  the displacement correlation function is $K(\omega) \propto 1/[\omega^2 - (\Omega + \rmi\gamma)^2]$, which yields
\begin{align}
K(t) = \overline{u^2}  \left( \cos \Omega t + \frac{\gamma}{\Omega}\sin \Omega |t|\right)\,\e^{-\gamma |t|} \,,
\end{align}
where $\Omega$ is the eigenfrequency of the oscillator, $\gamma$ is its damping rate, and $\overline{u^2}$ is the variance of the layer displacement. We suppose that $\gamma \ll \Omega, \Gamma'$ and obtain 
\begin{align}
C(\omega) =  \text{Re}\int_0^\infty \e^{k_0^2  \overline{u^2} (\cos \Omega t -1) + \rmi \omega t - \Gamma' t} dt \,.
\end{align}
Similarly to the case of periodic motion, the reflected pulse power spectra can be decomposed into a sum of Lorentzians 
 centered at the Raman sideband frequencies,
\begin{align}
\overline R(\omega) = \e^{-k_0^2  \overline{u^2} } \sum_{n=-\infty}^{\infty }I_n (k_0^2  \overline{u^2}\, ) \frac{\Gamma_0^2}{(\omega-n\Omega)^2 + \Gamma'^2} \,,
\end{align}
where $I_n$ is the modified Bessel function of the first kind. The same result can be obtained by averaging the result for periodic motion Eq.~\eqref{eq:Jn2} over the Gaussian distribution of motion amplitudes, $\int_0^\infty \overline R(\omega)\, u_0\,\e^{-u_0^2/(2\overline{u^2})}/\sqrt{\overline{u^2}} \,du_0$.

\begin{figure}[tb]
  \includegraphics[width=.99\columnwidth]{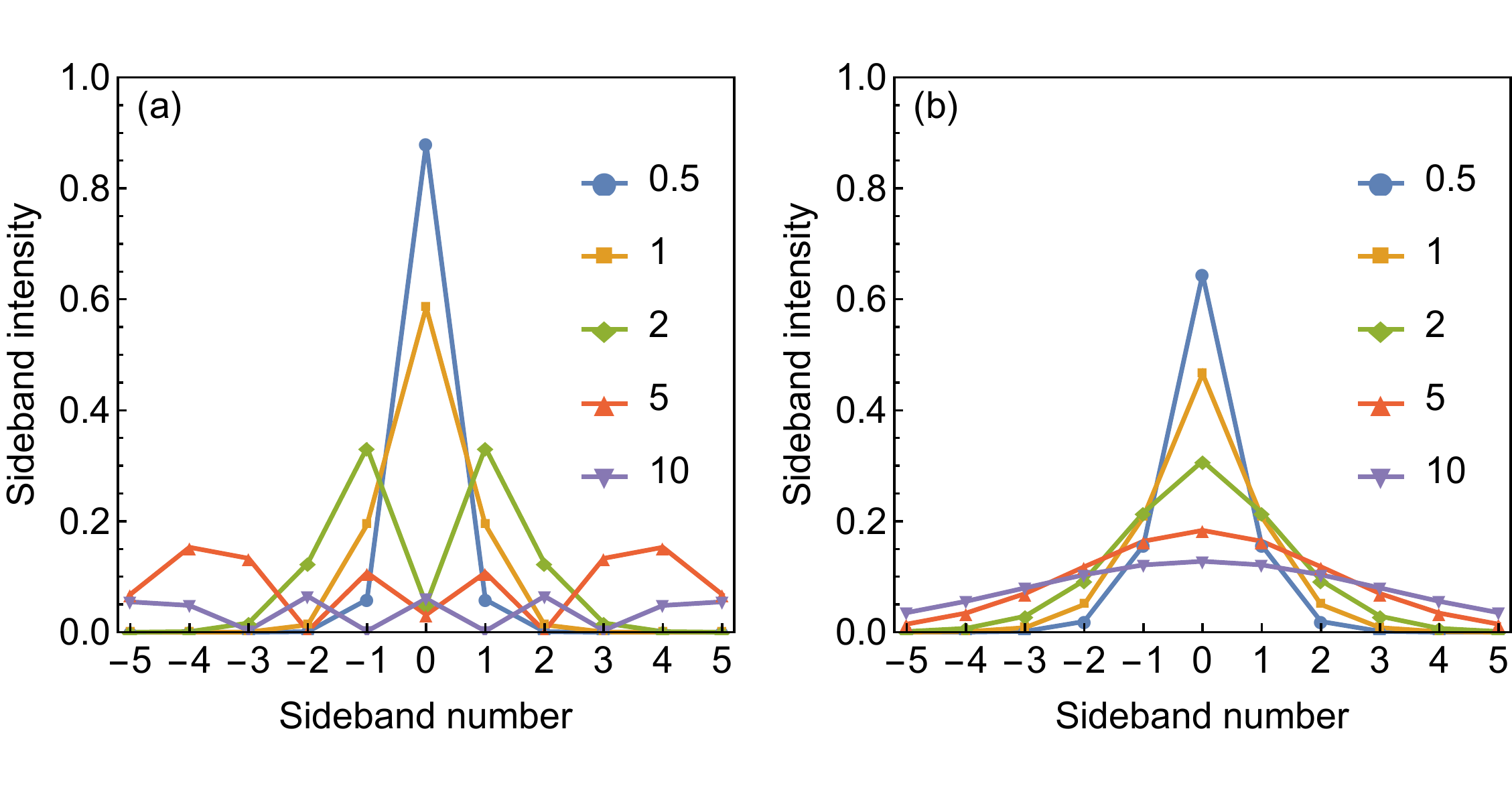}
\caption{The intensity of the Raman sidebands in the spectra of the pulse reflected from a resonant layer that moves (a) periodically as $u(t) = u_0 \sin \Omega t$ and (b) randomly with the correlation function $\langle u(0) u(t) \rangle = \overline{u^2} \cos \Omega t$.  Points correspond to different motion amplitudes, characterized by parameters $k_0u_0$ and $k_0\sqrt{\overline{u^2}}$, for the case of periodic and random motion, respectively. Lines are guide for eye. }\label{fig:side}
\end{figure}

Figure~\ref{fig:side} shows the intensities of the sidebands for different amplitudes of layer motion. Panels (a) and~(b) correspond to the cases of periodic and stochastic layer motion, respectively. For periodic harmonic layer motion, panel~(a) the sideband intensity oscillates with the sideband number. For large motion amplitudes, the most intensive sideband is that with the number $\propto k_0u_0$. 
In contrast, in the case of stochastic layer motion shown in panel~(b), the intensities of the  sidebands decay monotonously with the sideband number.

\section{Time-modulated resonance}
The essence of the considered Doppler-Raman crossover  is the periodic modulation of the layer resonant frequency due to Doppler shift. However, such modulation can be also realized directly. Generally, a deformation of the layer can lead to a shift of its resonance. This effect is especially strong in semiconductors, where the deformation potential mechanism leads to a strong shift of the exciton energy by $\delta\omega_x= \Xi \zeta$, where $\zeta$ is the deformation and $\Xi \sim 10\,\text{eV}$~\cite{Cardona,Demenev2019}.

Consider the quantum well with the exciton resonance frequency changing in time as
\begin{align}
\omega_r(t) = \omega_x + \delta\omega_x \cos \Omega t \,
\end{align}
due to the effect of an acoustic wave. 
The evolution of the exciton dipole polarization of the quantum well is described by the modified version of Eq.~\eqref{eq:evd} with $u_z=0$ and $\omega_0$ replaced with $\omega_r(t)$,
\begin{align}\label{eq:evd2}
\ddot{P_x} + \omega_r(t)^2 P_x + 2\Gamma'\dot P_x= - \frac{\Gamma_0}{\pi} \, \frac{\partial A_0}{\partial t}(0,t)   \,.
\end{align}
We again suppose that the frequency shifts are small $\delta\omega_x, \Omega \ll \omega_x$ and use the Wentzel--Kramers--Brillouin approximation to solve Eq.~\eqref{eq:evd2},
\begin{align}
P_x(t) = \frac{\Gamma_0 \theta (t-\tau)}{\pi\sqrt{\omega_r(t)\omega_r(\tau)}} \sin \left[ \int\limits^t_\tau \omega_r(t') dt'\right] \e^{-\Gamma' (t-\tau)} \,.
\end{align}
Then, the reflected light spectrum at the frequencies close to $\omega_x$ is given by
\begin{align}\label{eq:r2}
E_r(\omega)= -\Gamma_0 \int_0^{\infty} \e^{\rmi (\omega+\rmi\Gamma')t -\rmi \int_\tau^{\tau+t} \omega_r(t')dt' }\, dt \,
\end{align}
while the transmitted spectrum is $E_t(\omega) = 1 + E_r(\omega)$. In the absence of non-radiative damping, $\Gamma=0$, the conservation law has the form $\overline R(\omega) + \overline T(\omega) =1$.

The reflected pulse spectra for the cases of trembling layer and layer with oscillating resonance frequency, Eqs.~\eqref{eq:r} and~\eqref{eq:r2} match (up to a certain phase) if one takes $\delta\omega_x(t) = \omega_0 \dot{u}(t)$. That is exactly the condition of the Doppler shift for backscattered light being emulated by the direct modification of the resonance frequency. Similarly, the transmitted light spectra match if $\delta\omega_x(t) = -\omega_0 \dot{u}(t)$. 

\section{Conclusion}

To conclude, we have developed a theory of light interaction with moving resonant layers. Scattering of short pulses by layers oscillating in space was considered. Our calculation demonstrates  that when the motion frequency is small as compared to the resonance width, the spectrum of the reflected pulse features a peak at the frequency that changes in time in accordance with Doppler effect. In the opposite case of high layer motion frequency, the reflected spectrum comprises several Raman sidebands shifted by the multiples of mechanical frequency.  
The intensities of the sidebands are determined by the correlation function of layer displacement. 
Both effects, as well as the crossover between them, can be emulated  by use of the quantum well, where the exciton resonance frequency is modulated by strain.    

\section*{Acknowledgements}

This work was supported  by the Russian Science Foundation (project 20-42-04405). A.V.P. acknowledges the partial support from the Russian President Grant MK-599.2019.2 and the Foundation ``BASIS.''

\appendix
 
\section{Derivation of the action}\label{app:L}

Here, we derive  the Lagrangian describing the interaction between electromagnetic field and the moving dipole $\bm P(t)$. The dipole position at time $t$ is given by $\bm r= \bm u(t)$. 
We start from the action of a particle with charge $e$ in electromagnetic field,~\cite{landau02}
\begin{align}\label{eq:AL1}
S = e \int \bm A[\bm r(t), t] \dot{\bm r}(t) dt \,.
\end{align}
The dipole $\bm P(t)$ can be represented by two charges, $+e$ and $-e$, located at coordinates $\bm r = \bm u(t) + \bm P(t)/2e$ and $\bm r = \bm u(t) - \bm P(t)/2e$, respectively. Then, using Eq.~\eqref{eq:AL1} and considering the limit $ e \to \infty$ we obtain the action 
\begin{align}
S = &\int \left[(\bm P \cdot \bm\nabla) \bm A(\bm u, t) + \bm A( \bm u, t) \cdot \dot{\bm P}\right] dt \,.
\end{align}
Finally, we perform integration by parts and get
\begin{align}
S = \int \bm P \cdot \left[\bm E(\bm u, t) + \dot{\bm u} \times \bm B( \bm u, t) \right]   dt \,.
\end{align}
Note that here $\bm P$ is the dipole in the reference frame at rest. However, in the relevant case of $\bm u \perp \bm P$, it coincides with the dipole in the moving reference frame.


\begin{thebibliography}{17}%
\makeatletter
\providecommand \@ifxundefined [1]{%
 \@ifx{#1\undefined}
}%
\providecommand \@ifnum [1]{%
 \ifnum #1\expandafter \@firstoftwo
 \else \expandafter \@secondoftwo
 \fi
}%
\providecommand \@ifx [1]{%
 \ifx #1\expandafter \@firstoftwo
 \else \expandafter \@secondoftwo
 \fi
}%
\providecommand \natexlab [1]{#1}%
\providecommand \enquote  [1]{``#1''}%
\providecommand \bibnamefont  [1]{#1}%
\providecommand \bibfnamefont [1]{#1}%
\providecommand \citenamefont [1]{#1}%
\providecommand \href@noop [0]{\@secondoftwo}%
\providecommand \href [0]{\begingroup \@sanitize@url \@href}%
\providecommand \@href[1]{\@@startlink{#1}\@@href}%
\providecommand \@@href[1]{\endgroup#1\@@endlink}%
\providecommand \@sanitize@url [0]{\catcode `\\12\catcode `\$12\catcode
  `\&12\catcode `\#12\catcode `\^12\catcode `\_12\catcode `\%12\relax}%
\providecommand \@@startlink[1]{}%
\providecommand \@@endlink[0]{}%
\providecommand \url  [0]{\begingroup\@sanitize@url \@url }%
\providecommand \@url [1]{\endgroup\@href {#1}{\urlprefix }}%
\providecommand \urlprefix  [0]{URL }%
\providecommand \Eprint [0]{\href }%
\providecommand \doibase [0]{http://dx.doi.org/}%
\providecommand \selectlanguage [0]{\@gobble}%
\providecommand \bibinfo  [0]{\@secondoftwo}%
\providecommand \bibfield  [0]{\@secondoftwo}%
\providecommand \translation [1]{[#1]}%
\providecommand \BibitemOpen [0]{}%
\providecommand \bibitemStop [0]{}%
\providecommand \bibitemNoStop [0]{.\EOS\space}%
\providecommand \EOS [0]{\spacefactor3000\relax}%
\providecommand \BibitemShut  [1]{\csname bibitem#1\endcsname}%
\let\auto@bib@innerbib\@empty
\bibitem [{\citenamefont {Boyd}(2003)}]{boyd2003nonlinear}%
  \BibitemOpen
  \bibfield  {author} {\bibinfo {author} {\bibfnamefont {R.~W.}\ \bibnamefont
  {Boyd}},\ }\href@noop {} {\emph {\bibinfo {title} {Nonlinear optics}}}\
  (\bibinfo  {publisher} {Academic press},\ \bibinfo {year} {2003})\BibitemShut
  {NoStop}%
\bibitem [{\citenamefont {Fan}\ \emph {et~al.}(2016)\citenamefont {Fan},
  \citenamefont {Zou}, \citenamefont {Poot}, \citenamefont {Cheng},
  \citenamefont {Guo}, \citenamefont {Han},\ and\ \citenamefont
  {Tang}}]{Fan2016}%
  \BibitemOpen
  \bibfield  {author} {\bibinfo {author} {\bibfnamefont {L.}~\bibnamefont
  {Fan}}, \bibinfo {author} {\bibfnamefont {C.-L.}\ \bibnamefont {Zou}},
  \bibinfo {author} {\bibfnamefont {M.}~\bibnamefont {Poot}}, \bibinfo {author}
  {\bibfnamefont {R.}~\bibnamefont {Cheng}}, \bibinfo {author} {\bibfnamefont
  {X.}~\bibnamefont {Guo}}, \bibinfo {author} {\bibfnamefont {X.}~\bibnamefont
  {Han}}, \ and\ \bibinfo {author} {\bibfnamefont {H.~X.}\ \bibnamefont
  {Tang}},\ }\href {\doibase 10.1038/nphoton.2016.206} {\bibfield  {journal}
  {\bibinfo  {journal} {Nature Photonics}\ }\textbf {\bibinfo {volume} {10}},\
  \bibinfo {pages} {766} (\bibinfo {year} {2016})}\BibitemShut {NoStop}%
\bibitem [{\citenamefont {Preble}\ \emph {et~al.}(2007)\citenamefont {Preble},
  \citenamefont {Xu},\ and\ \citenamefont {Lipson}}]{Preble2007}%
  \BibitemOpen
  \bibfield  {author} {\bibinfo {author} {\bibfnamefont {S.~F.}\ \bibnamefont
  {Preble}}, \bibinfo {author} {\bibfnamefont {Q.}~\bibnamefont {Xu}}, \ and\
  \bibinfo {author} {\bibfnamefont {M.}~\bibnamefont {Lipson}},\ }\href
  {\doibase 10.1038/nphoton.2007.72} {\bibfield  {journal} {\bibinfo  {journal}
  {Nature Photonics}\ }\textbf {\bibinfo {volume} {1}},\ \bibinfo {pages} {293}
  (\bibinfo {year} {2007})}\BibitemShut {NoStop}%
\bibitem [{\citenamefont {Scruby}\ and\ \citenamefont
  {Drain}(1990)}]{scruby_laser_1990}%
  \BibitemOpen
  \bibfield  {author} {\bibinfo {author} {\bibfnamefont {C.~B.}\ \bibnamefont
  {Scruby}}\ and\ \bibinfo {author} {\bibfnamefont {L.~E.}\ \bibnamefont
  {Drain}},\ }\href@noop {} {\emph {\bibinfo {title} {Laser ultrasonics:
  techniques and applications}}}\ (\bibinfo  {publisher} {A. Hilger},\ \bibinfo
  {address} {Bristol, England ; Philadelphia},\ \bibinfo {year}
  {1990})\BibitemShut {NoStop}%
\bibitem [{\citenamefont {Kuramochi}\ and\ \citenamefont
  {Notomi}(2016)}]{Kuramochi2016}%
  \BibitemOpen
  \bibfield  {author} {\bibinfo {author} {\bibfnamefont {E.}~\bibnamefont
  {Kuramochi}}\ and\ \bibinfo {author} {\bibfnamefont {M.}~\bibnamefont
  {Notomi}},\ }\href {\doibase 10.1038/nphoton.2016.238} {\bibfield  {journal}
  {\bibinfo  {journal} {Nature Photonics}\ }\textbf {\bibinfo {volume} {10}},\
  \bibinfo {pages} {752} (\bibinfo {year} {2016})}\BibitemShut {NoStop}%
\bibitem [{\citenamefont {Demenev}\ \emph {et~al.}(2019)\citenamefont
  {Demenev}, \citenamefont {Yaremkevich}, \citenamefont {Scherbakov},
  \citenamefont {Kukhtaruk}, \citenamefont {Gavrilov}, \citenamefont
  {Yakovlev}, \citenamefont {Kulakovskii},\ and\ \citenamefont
  {Bayer}}]{Demenev2019}%
  \BibitemOpen
  \bibfield  {author} {\bibinfo {author} {\bibfnamefont {A.~A.}\ \bibnamefont
  {Demenev}}, \bibinfo {author} {\bibfnamefont {D.~D.}\ \bibnamefont
  {Yaremkevich}}, \bibinfo {author} {\bibfnamefont {A.~V.}\ \bibnamefont
  {Scherbakov}}, \bibinfo {author} {\bibfnamefont {S.~M.}\ \bibnamefont
  {Kukhtaruk}}, \bibinfo {author} {\bibfnamefont {S.~S.}\ \bibnamefont
  {Gavrilov}}, \bibinfo {author} {\bibfnamefont {D.~R.}\ \bibnamefont
  {Yakovlev}}, \bibinfo {author} {\bibfnamefont {V.~D.}\ \bibnamefont
  {Kulakovskii}}, \ and\ \bibinfo {author} {\bibfnamefont {M.}~\bibnamefont
  {Bayer}},\ }\href {\doibase 10.1103/PhysRevB.100.100301} {\bibfield
  {journal} {\bibinfo  {journal} {Phys. Rev. B}\ }\textbf {\bibinfo {volume}
  {100}},\ \bibinfo {pages} {100301} (\bibinfo {year} {2019})}\BibitemShut
  {NoStop}%
\bibitem [{\citenamefont {Farmer}\ \emph {et~al.}(2014)\citenamefont {Farmer},
  \citenamefont {Akimov}, \citenamefont {Gippius}, \citenamefont {Bailey},
  \citenamefont {Sharp},\ and\ \citenamefont {Kent}}]{Farmer2014}%
  \BibitemOpen
  \bibfield  {author} {\bibinfo {author} {\bibfnamefont {D.~J.}\ \bibnamefont
  {Farmer}}, \bibinfo {author} {\bibfnamefont {A.~V.}\ \bibnamefont {Akimov}},
  \bibinfo {author} {\bibfnamefont {N.~A.}\ \bibnamefont {Gippius}}, \bibinfo
  {author} {\bibfnamefont {J.}~\bibnamefont {Bailey}}, \bibinfo {author}
  {\bibfnamefont {J.~S.}\ \bibnamefont {Sharp}}, \ and\ \bibinfo {author}
  {\bibfnamefont {A.~J.}\ \bibnamefont {Kent}},\ }\href {\doibase
  10.1364/oe.22.015218} {\bibfield  {journal} {\bibinfo  {journal} {Optics
  Express}\ }\textbf {\bibinfo {volume} {22}},\ \bibinfo {pages} {15218}
  (\bibinfo {year} {2014})}\BibitemShut {NoStop}%
\bibitem [{\citenamefont {Jusserand}\ \emph {et~al.}(2015)\citenamefont
  {Jusserand}, \citenamefont {Poddubny}, \citenamefont {Poshakinskiy},
  \citenamefont {Fainstein},\ and\ \citenamefont {Lemaitre}}]{Jusserand2015}%
  \BibitemOpen
  \bibfield  {author} {\bibinfo {author} {\bibfnamefont {B.}~\bibnamefont
  {Jusserand}}, \bibinfo {author} {\bibfnamefont {A.~N.}\ \bibnamefont
  {Poddubny}}, \bibinfo {author} {\bibfnamefont {A.~V.}\ \bibnamefont
  {Poshakinskiy}}, \bibinfo {author} {\bibfnamefont {A.}~\bibnamefont
  {Fainstein}}, \ and\ \bibinfo {author} {\bibfnamefont {A.}~\bibnamefont
  {Lemaitre}},\ }\href {\doibase 10.1103/PhysRevLett.115.267402} {\bibfield
  {journal} {\bibinfo  {journal} {Phys. Rev. Lett.}\ }\textbf {\bibinfo
  {volume} {115}},\ \bibinfo {pages} {267402} (\bibinfo {year}
  {2015})}\BibitemShut {NoStop}%
\bibitem [{\citenamefont {Br\"{u}ggemann}\ \emph {et~al.}(2011)\citenamefont
  {Br\"{u}ggemann}, \citenamefont {Akimov}, \citenamefont {Scherbakov},
  \citenamefont {Bombeck}, \citenamefont {Schneider}, \citenamefont
  {H\"{o}fling}, \citenamefont {Forchel}, \citenamefont {Yakovlev},\ and\
  \citenamefont {Bayer}}]{Brggemann2011}%
  \BibitemOpen
  \bibfield  {author} {\bibinfo {author} {\bibfnamefont {C.}~\bibnamefont
  {Br\"{u}ggemann}}, \bibinfo {author} {\bibfnamefont {A.~V.}\ \bibnamefont
  {Akimov}}, \bibinfo {author} {\bibfnamefont {A.~V.}\ \bibnamefont
  {Scherbakov}}, \bibinfo {author} {\bibfnamefont {M.}~\bibnamefont {Bombeck}},
  \bibinfo {author} {\bibfnamefont {C.}~\bibnamefont {Schneider}}, \bibinfo
  {author} {\bibfnamefont {S.}~\bibnamefont {H\"{o}fling}}, \bibinfo {author}
  {\bibfnamefont {A.}~\bibnamefont {Forchel}}, \bibinfo {author} {\bibfnamefont
  {D.~R.}\ \bibnamefont {Yakovlev}}, \ and\ \bibinfo {author} {\bibfnamefont
  {M.}~\bibnamefont {Bayer}},\ }\href {\doibase 10.1038/nphoton.2011.269}
  {\bibfield  {journal} {\bibinfo  {journal} {Nat. Photonics}\ }\textbf
  {\bibinfo {volume} {6}},\ \bibinfo {pages} {30} (\bibinfo {year}
  {2011})}\BibitemShut {NoStop}%
\bibitem [{\citenamefont {Berstermann}\ \emph {et~al.}(2009)\citenamefont
  {Berstermann}, \citenamefont {Scherbakov}, \citenamefont {Akimov},
  \citenamefont {Yakovlev}, \citenamefont {Gippius}, \citenamefont {Glavin},
  \citenamefont {Sagnes}, \citenamefont {Bloch},\ and\ \citenamefont
  {Bayer}}]{Berstermann2009}%
  \BibitemOpen
  \bibfield  {author} {\bibinfo {author} {\bibfnamefont {T.}~\bibnamefont
  {Berstermann}}, \bibinfo {author} {\bibfnamefont {A.~V.}\ \bibnamefont
  {Scherbakov}}, \bibinfo {author} {\bibfnamefont {A.~V.}\ \bibnamefont
  {Akimov}}, \bibinfo {author} {\bibfnamefont {D.~R.}\ \bibnamefont
  {Yakovlev}}, \bibinfo {author} {\bibfnamefont {N.~A.}\ \bibnamefont
  {Gippius}}, \bibinfo {author} {\bibfnamefont {B.~A.}\ \bibnamefont {Glavin}},
  \bibinfo {author} {\bibfnamefont {I.}~\bibnamefont {Sagnes}}, \bibinfo
  {author} {\bibfnamefont {J.}~\bibnamefont {Bloch}}, \ and\ \bibinfo {author}
  {\bibfnamefont {M.}~\bibnamefont {Bayer}},\ }\href {\doibase
  10.1103/PhysRevB.80.075301} {\bibfield  {journal} {\bibinfo  {journal} {Phys.
  Rev. B}\ }\textbf {\bibinfo {volume} {80}},\ \bibinfo {pages} {075301}
  (\bibinfo {year} {2009})}\BibitemShut {NoStop}%
\bibitem [{\citenamefont {Berstermann}\ \emph {et~al.}(2012)\citenamefont
  {Berstermann}, \citenamefont {Br\"uggemann}, \citenamefont {Akimov},
  \citenamefont {Bombeck}, \citenamefont {Yakovlev}, \citenamefont {Gippius},
  \citenamefont {Scherbakov}, \citenamefont {Sagnes}, \citenamefont {Bloch},\
  and\ \citenamefont {Bayer}}]{Berstermann2012}%
  \BibitemOpen
  \bibfield  {author} {\bibinfo {author} {\bibfnamefont {T.}~\bibnamefont
  {Berstermann}}, \bibinfo {author} {\bibfnamefont {C.}~\bibnamefont
  {Br\"uggemann}}, \bibinfo {author} {\bibfnamefont {A.~V.}\ \bibnamefont
  {Akimov}}, \bibinfo {author} {\bibfnamefont {M.}~\bibnamefont {Bombeck}},
  \bibinfo {author} {\bibfnamefont {D.~R.}\ \bibnamefont {Yakovlev}}, \bibinfo
  {author} {\bibfnamefont {N.~A.}\ \bibnamefont {Gippius}}, \bibinfo {author}
  {\bibfnamefont {A.~V.}\ \bibnamefont {Scherbakov}}, \bibinfo {author}
  {\bibfnamefont {I.}~\bibnamefont {Sagnes}}, \bibinfo {author} {\bibfnamefont
  {J.}~\bibnamefont {Bloch}}, \ and\ \bibinfo {author} {\bibfnamefont
  {M.}~\bibnamefont {Bayer}},\ }\href {\doibase 10.1103/PhysRevB.86.195306}
  {\bibfield  {journal} {\bibinfo  {journal} {Phys. Rev. B}\ }\textbf {\bibinfo
  {volume} {86}},\ \bibinfo {pages} {195306} (\bibinfo {year}
  {2012})}\BibitemShut {NoStop}%
\bibitem [{\citenamefont {Poshakinskiy}\ and\ \citenamefont
  {Poddubny}(2019)}]{Poshakinskiy2019}%
  \BibitemOpen
  \bibfield  {author} {\bibinfo {author} {\bibfnamefont {A.~V.}\ \bibnamefont
  {Poshakinskiy}}\ and\ \bibinfo {author} {\bibfnamefont {A.~N.}\ \bibnamefont
  {Poddubny}},\ }\href {\doibase 10.1103/PhysRevX.9.011008} {\bibfield
  {journal} {\bibinfo  {journal} {Phys. Rev. X}\ }\textbf {\bibinfo {volume}
  {9}},\ \bibinfo {pages} {011008} (\bibinfo {year} {2019})}\BibitemShut
  {NoStop}%
\bibitem [{\citenamefont {Andreani}\ \emph {et~al.}(1998)\citenamefont
  {Andreani}, \citenamefont {Panzarini}, \citenamefont {Kavokin},\ and\
  \citenamefont {Vladimirova}}]{Andreani1998}%
  \BibitemOpen
  \bibfield  {author} {\bibinfo {author} {\bibfnamefont {L.~C.}\ \bibnamefont
  {Andreani}}, \bibinfo {author} {\bibfnamefont {G.}~\bibnamefont {Panzarini}},
  \bibinfo {author} {\bibfnamefont {A.~V.}\ \bibnamefont {Kavokin}}, \ and\
  \bibinfo {author} {\bibfnamefont {M.~R.}\ \bibnamefont {Vladimirova}},\
  }\href {\doibase 10.1103/PhysRevB.57.4670} {\bibfield  {journal} {\bibinfo
  {journal} {Phys. Rev. B}\ }\textbf {\bibinfo {volume} {57}},\ \bibinfo
  {pages} {4670} (\bibinfo {year} {1998})}\BibitemShut {NoStop}%
\bibitem [{\citenamefont {Ammerlahn}\ \emph {et~al.}(2000)\citenamefont
  {Ammerlahn}, \citenamefont {Kuhl}, \citenamefont {Grote}, \citenamefont
  {Koch}, \citenamefont {Khitrova},\ and\ \citenamefont
  {Gibbs}}]{Ammerlahn2000}%
  \BibitemOpen
  \bibfield  {author} {\bibinfo {author} {\bibfnamefont {D.}~\bibnamefont
  {Ammerlahn}}, \bibinfo {author} {\bibfnamefont {J.}~\bibnamefont {Kuhl}},
  \bibinfo {author} {\bibfnamefont {B.}~\bibnamefont {Grote}}, \bibinfo
  {author} {\bibfnamefont {S.~W.}\ \bibnamefont {Koch}}, \bibinfo {author}
  {\bibfnamefont {G.}~\bibnamefont {Khitrova}}, \ and\ \bibinfo {author}
  {\bibfnamefont {H.}~\bibnamefont {Gibbs}},\ }\href {\doibase
  10.1103/PhysRevB.62.7350} {\bibfield  {journal} {\bibinfo  {journal} {Phys.
  Rev. B}\ }\textbf {\bibinfo {volume} {62}},\ \bibinfo {pages} {7350}
  (\bibinfo {year} {2000})}\BibitemShut {NoStop}%
\bibitem [{\citenamefont {Poshakinskiy}\ \emph {et~al.}(2012)\citenamefont
  {Poshakinskiy}, \citenamefont {Poddubny},\ and\ \citenamefont
  {Tarasenko}}]{Poshakinskiy2012}%
  \BibitemOpen
  \bibfield  {author} {\bibinfo {author} {\bibfnamefont {A.~V.}\ \bibnamefont
  {Poshakinskiy}}, \bibinfo {author} {\bibfnamefont {A.~N.}\ \bibnamefont
  {Poddubny}}, \ and\ \bibinfo {author} {\bibfnamefont {S.~A.}\ \bibnamefont
  {Tarasenko}},\ }\href {\doibase 10.1103/PhysRevB.86.205304} {\bibfield
  {journal} {\bibinfo  {journal} {Phys. Rev. B}\ }\textbf {\bibinfo {volume}
  {86}},\ \bibinfo {pages} {205304} (\bibinfo {year} {2012})}\BibitemShut
  {NoStop}%
\bibitem [{\citenamefont {Yu}\ and\ \citenamefont {Cardona}(2010)}]{Cardona}%
  \BibitemOpen
  \bibfield  {author} {\bibinfo {author} {\bibfnamefont {P.}~\bibnamefont
  {Yu}}\ and\ \bibinfo {author} {\bibfnamefont {M.}~\bibnamefont {Cardona}},\
  }\href {http://books.google.se/books?id=5aBuKYBT\_hsC} {\emph {\bibinfo
  {title} {{F}undamentals of {S}emiconductors: {P}hysics and {M}aterials
  {P}roperties}}},\ Graduate texts in physics\ (\bibinfo  {publisher}
  {Springer},\ \bibinfo {year} {2010})\BibitemShut {NoStop}%
\bibitem [{\citenamefont {Landau}\ and\ \citenamefont
  {Lifshitz}(2009)}]{landau02}%
  \BibitemOpen
  \bibfield  {author} {\bibinfo {author} {\bibfnamefont {L.}~\bibnamefont
  {Landau}}\ and\ \bibinfo {author} {\bibfnamefont {E.}~\bibnamefont
  {Lifshitz}},\ }\href@noop {} {\emph {\bibinfo {title} {The classical theory
  of fields}}},\ \bibinfo {edition} {4th}\ ed.,\ \bibinfo {series} {Course of
  theoretical physics}\ No.~\bibinfo {number} {2}\ (\bibinfo  {publisher}
  {Elsevier},\ \bibinfo {address} {Amsterdam},\ \bibinfo {year}
  {2009})\BibitemShut {NoStop}%
\end{thebibliography}

%

\end{document}